# Dynamics of reallocation within India's income distribution

Anand Sahasranaman[1,2,*] and Henrik Jeldtoft Jensen[2,3#]

[1]Division of Science, Krea University, Andhra Pradesh 517646, India.

[2]Centre for Complexity Science, Dept of Mathematics, Imperial College London, London SW72AZ, UK.

[3]Institute of Innovative Research, Tokyo Institute of Technology, 4259, Nagatsuta-cho, Yokohama 226-8502, Japan

[*] Corresponding Author. Email: anand.sahasranaman@krea.edu.in

[#] Email: h.jensen@imperial.ac.uk

**Abstract:**

We investigate the nature and extent of reallocation occurring within the Indian income distribution, with a particular focus on the dynamics of the bottom of the distribution. Specifically, we use a stochastic model of Geometric Brownian Motion with a reallocation parameter that was constructed to capture the quantum and direction of composite redistribution implied in the income distribution. It is well known that inequality has been rising in India in the recent past, but the assumption has been that while the rich benefit more than proportionally from economic growth, the poor are also better off than before. Findings from our model refute this, as we find that since the early 2000s reallocation has consistently been negative, and that the Indian income distribution has entered a regime of perverse redistribution of resources from the poor to the rich. Outcomes from the model indicate not only that income shares of the bottom decile (~1%) and bottom percentile (~0.03%) are at historic lows, but also that real incomes of the bottom decile (-2.5%) and percentile (-6%) have declined in the 2000s. We validate these findings using income distribution data and find support for our contention of persistent negative reallocation in the 2000s. We characterize these findings in the context of increasing informalization of the workforce in the formal manufacturing and service sectors, as well as the growing economic insecurity of the agricultural workforce in India. Significant structural changes will be required to address this phenomenon.

# 1. Introduction

We live in a time characterized by increasing anxiety about economic inequality (Ribeiro, 2013; Oncu, 2013; Lyster, 2016; Kohut, 2011). Since the early 1980s, there has been a systematic growth of income inequality in nations across the world, and India has been no exception (Milanovic, 2016). And while significant attention has focused on India's poverty alleviation effects over the past 40 years (World Bank, 2019; Dhongde, 2007; Ninan, 1994; Dev & Ravi, 2007; Deaton & Dreze, 2002; Kjelsrud & Somanathan, 2017), the rapid rise in inequality in the same period merits deeper examination, especially pertaining to the dynamics at the bottom of the distribution. Many studies on the Indian income distribution use consumption data as provided by the National Sample Survey (NSS) as the basis to study income inequality (Deaton & Dreze, 2002; Sarkar & Mehta, 2010), while others have attempted to construct the income distribution using multiple sources in addition to NSS expenditure data, such as income tax data, national accounts data from the Central Statistical Organization (CSO), and Reserve Bank of India's (RBI) household savings data (Ojha & Bhatt, 1964; Ahmed & Bhattacharya, 2017; Sinha, Pearson, Kadekodi, & Gregory, 2017; Banerjee & Piketty, 2005; Chancel & Piketty, 2019). Chancel and Piketty (2019), construct the longest (and most up-to-date) income inequality time series for India, from 1922 to 2015, using income tax data, NSS expenditure data, and the India Human Development Survey (IHDS) income data, providing us a picture of the longer-term temporal evolution of income inequality in India (Figure1). Income inequality shows a declining trend for the first three decades after independence, with the top 10% (1%) earning 36.7% (11.5%) of the total income in 1951 and 30.7% (6.7%) in 1981. The bottom 50% meanwhile see their income share increase from 20.6% to 23.5% in the same period. However, from the early 1980s, inequality has shown a sustained and steep increase, resulting in the top 10% (1%) earning 56% (21%) of income in 2015, with the bottom 50% seeing their share reduce to 14.7% (Chancel & Piketty, 2019).

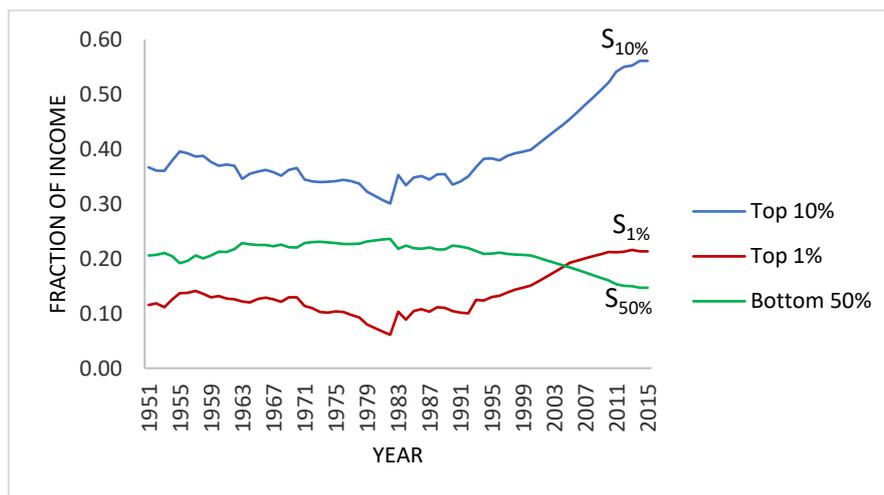

**Figure 1:** *Evolution of income inequality (1951 – 2015)*: The temporal evolution of inequality is represented through the shares of incomes owned by the top 1% ($S_{1\%}$, red line), top 10% ($S_{10\%}$, blue line), and bottom 50% ($S_{50\%}$, green line) of the population. After top income shares declined until the early 1980s, they have sharply risen since then. Income shares of the bottom half have declined.

Before we delve deeper into the dynamics of inequality in India, it is useful to contextualize the Indian experience within the broader global experience of inequality. Prior to the

Industrial Revolution, mean incomes in most countries were stagnant for many centuries (Alvarez-Nogal & Prados de la Escosura, 2013), but inequality waxed and waned over time as a consequence of idiosyncratic forces such as wars, discovery of new lands, and epidemics (Milanovic, 2016). The rise and fall of inequality around an essentially fixed mean income illustrates the fact that there was no systematic relationship between inequality and income. The industrial revolution however, appears to have fundamentally altered the dynamic between income and inequality in two significant ways (Milanovic, 2016). First, growing total national incomes meant that inequality had more 'space' to increase now than before, thereby allowing a small portion of the population very high incomes, while also ensuring that nobody was pushed below subsistence level. This notion of greater potential inequality on account of increasing total income has been formalized as the 'inequality possibility frontier', which is defined as the locus of maximum feasible inequality levels for different values of mean income (Milanovic, Lindert, & Williamson, 2011). Second, after the Industrial Revolution there emerged a new relationship between mean income and inequality. Both mean income and income inequality, on average, displayed a rising trend over time. The structural change on account of shift in occupations from agriculture to industry as well as changes in patterns of living as captured in the rural to urban migration, drove inequality up as a consequence of capital being able to capture most of the gains of increasing total income at the expense of labour. It has been argued that income inequality always rises when the rate of return from capital is greater than the rate of economic growth (Piketty, 2014), and that it is only for the brief period in the middle of the 20$^{th}$ century that there is a decline in inequality, which is due to a special set of political circumstances such as education, taxation, workers movements, social security, as well as economic convergence (Milanovic, 2016; Piketty, 2014). This decline is apparent in the time evolution of inequality between the second world war and the 1980s - to illustrate, between 1955 and 1980, the share of income earned by the top 10% declined by 7% in the United States, 15% in France, 18% in the Soviet Union, and 19% in India (Alvaredo, Chancel, Piketty, Saez, & Zucman, 2017; Chancel & Piketty, 2019; Novokmet, Piketty, & Zucman, 2017; Garbinti, Goupille-Lebret, & Piketty, 2017). However, post the 1980s, inequality has resumed its expected upward trend and Milanovic (2016) argues that we are currently witnessing yet another set of structural changes encompassing the communications and internet revolution that has resulted in a sectoral shift from industry to services, increased economic interconnectedness between countries, and weakened the labour movement on account of the dispersed nature of employment in the services industry. Again, capital has captured a large share of the increased total income, resulting in a rising trend of economic inequality in nations across the world, even as average incomes have continued rising. For instance, between 1995 and 2012 the total fraction of income earned by the top 10% has grown by almost 13% in South Africa, 15% in the US, 25% in China, and 45% in India (Alvaredo, Chancel, Piketty, Saez, & Zucman, 2017; Chancel & Piketty, 2019; Assouad, Chancel, & Morgan, 2018).

While the evolution of income inequality in India appears to broadly follow global trends, it is important to recognize that constructing the income distribution for measurement of inequality in India presents specific and unique challenges. In order to construct an annual time series of income inequality for India, given the largely informal nature of the workforce (tax data only covers ~7% of the working population), incomes of over 90% of the population are generally estimated from NSS consumption data because regular income surveys do not exist (Chancel & Piketty, 2019; Bardhan, 2017). There are a number of challenges, such as under-reporting and under-sampling, in using NSS consumption data to estimate income inequality, in addition to the fact that the dynamics of these two distributions (consumption and income) may be quite different (Atkinson & Piketty, 2010). In their work estimating India's income distribution from 1922 to 2015, Chancel and Piketty (2019) rely on tax data for the top of the distribution, but use income data from two IHDS surveys to compute income-consumption ratios, which forms their basis to construct income profiles from NSS consumption data. They clearly discuss the significant methodological and data challenges inherent in the empirical construction of the entire Indian income distribution, especially moving towards the bottom incomes.

Studies of inequality generally tend to focus on the income shares of those at the top of the distribution (top 0.1%, top 1%, top 10% etc.), so as to understand the (often disproportionate) extent of economic growth they have garnered over time. Our primary interest, however, lies in understanding the nature and extent of reallocation occurring within the income distribution over time. We propose to use income inequality data to fit a stochastic model of income evolution and thereby construct a theoretically consistent estimation of redistribution inherent in the economy. We also seek to understand better the dynamics of the lowest end of the spectrum – the bottom decile and the bottom percentile of the income distribution. Finally, we discuss the implications of our findings in the context of significant global and national trends impinging on the Indian economy.

## 2. Model definition and specifications

In order to contemplate appropriate models to simulate income growth over time, it is useful to go back to the systematic nature of the relationship between mean income and income inequality post the Industrial revolution (Milanovic, 2016; Piketty, 2014) – both quantities, on average, are found to rise over time. Given this framing of income dynamics, income evolution is well suited to be studied as a multiplicative growth process following Geometric Brownian Motion (GBM), which obtains a broadening lognormal distribution over time. Indeed, empirical studies of income distributions around the world suggest that multiplicative dynamics yielding exponential or log normal distributions are salient for the lower part of distributions, with power laws operational at the tails of the distribution (Banerjee, Yakovenko, & Di Matteo, 2006; Clementi & Gallegati, 2005; Drăgulescu & Yakovenko, 2001; Souma, 2001). Beside income, many economic processes such as evolution of wealth and asset prices have been modelled as multiplicative processes (Bouchaud & Mezard, 2000; Vasicek, 1977; Berman, Peters, & Adamou, 2017; Gabaix, Lasry, Lions, & Moll, 2016).

Using NSS consumption data for India (given the absence of income time series in India as discussed earlier), it was found that the distribution of consumption expenditures showed a lognormal body and a power law tail (Chatterjee, Chakrabarti, Ghosh, Chakraborti, & Nandi, 2016; Ghosh, Gangopadhyay, & Basu, 2011). Also, when we study the evolution of mean (per capita) national income from 1947 to 2017, it is found to be reasonably approximated by an exponential function (Figure 2).

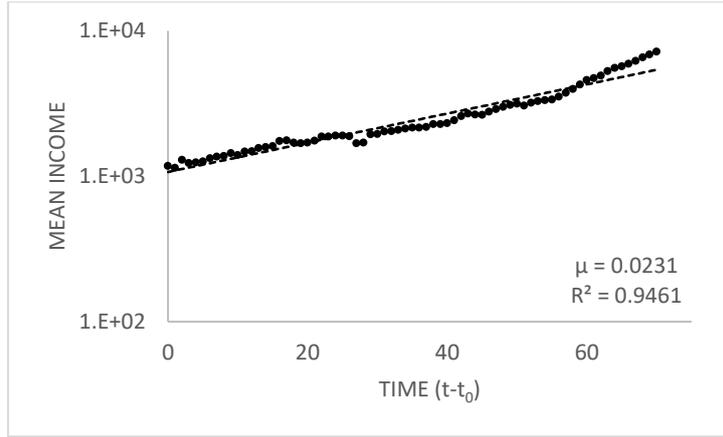

**Figure 2:** *Evolution of mean per capita income (1947 – 2017)*: The temporal evolution of mean income (black dots, log scale) is well approximated by an exponential process (dashed black line) of the form $exp[\mu(t - t_0)]$, yielding $\mu = 0.0231$. Time $t_0 = 0$ represents 1947 and each subsequent unit time increment represents one additional year, until 2017.

In this work, we propose to propagate individual incomes using a multiplicative growth process, with the objective of estimating the direction and quantum of redistribution occurring in the income distribution. Specifically, we use the Reallocating GBM (RGBM) methodology of Berman, Peters, and Adamou (2017), who analyse wealth dynamics under disequilibrium – i.e. without the assumption that rescaled wealth converges to a stationary distribution. Essentially, using the RGBM approach, we model income as a noisy multiplicative process following GBM, while incorporating a reallocation parameter ($\tau$) to capture the transfer of income between individuals.

The reallocation parameter in this model is a measure of overall reallocation occurring in the income distribution. If we conceptualize individual incomes as a cumulative outcome composed of both systemic inputs such as public investments, economic environment, labour regulations, and tax laws, as well as individual idiosyncratic inputs, then the reallocation parameter ($\tau$) is best understood as a measure encapsulating the consolidated redistributive impact of all these factors as manifested in the resultant income distribution.

Under the RGBM, the time-evolution of income comprises two mechanisms, namely growth and reallocation, and is modelled using the following stochastic differential equation (Berman, Peters, & Adamou, 2017) (Eq.1):

$$dx_i = x_i(\mu dt + \sigma dW_i) - \tau(x_i - \langle x \rangle_N), \qquad (1)$$

where:

$$\langle x \rangle_N = \frac{1}{N}\sum_{i=1}^{N} x_i \qquad (2)$$

$dx_i$ is the change in income of individual $i$ over time period $dt$. The first term $x_i(\mu dt + \sigma dW_i)$ is the growth term and the second one $\tau(x_i - \langle x \rangle_N)dt$ is the reallocation term. In the growth term, the $\mu dt$ represents systemic growth (economic growth that affects all incomes), while the $\sigma dW_i$ represents idiosyncratic growth of the particular individual $i$'s income, with $dW_i$ specifically being the increment in a Wiener process, which is normally distributed with mean zero and variance $dt$. $x_i$ is the income of $i$ at time $t$, while the parameters $\mu$ and $\sigma$ stand for drift and volatility of income respectively. The reallocation term comprises the reallocation parameter $\tau$ applied to the net reallocation from individual $i$, which is the difference between the individual's income $x_i$ and mean income $\langle x \rangle_N$.

Income inequality time series data for India (1922-2017) is obtained from the World Inequality Database (https://wid.world/country/india/) (Chancel & Piketty, 2019). We earlier highlighted the challenges inherent in their construction of the income time series for India, but despite these concerns, Chancel and Piketty (2019) find that their key, coarse-grained results on the temporal evolution of income shares of the top 1%, top 10%, and bottom 50% are robust to a range of assumptions. We use these robust inequality measures of Chancel and Piketty (2019) to fit the RGBM model and estimate reallocation within the distribution. We use this data for the period from 1951, when India became a republic, until 2015. Specifically, the dataset provides annual estimates on incomes of the top 1%, 10% and bottom 50% of population as proportions of total national income – $S_{1\%}, S_{10\%}, S_{50\%}$ respectively (Figure1). Our interest is in exploring the dynamics at the lower (poor) end of the distribution, and given that the (rich) Pareto tail of the distribution could possibly cover between 10% and 20% of the population (Ghosh, Gangopadhyay, & Basu, 2011), using either the $S_{1\%}$ or $S_{10\%}$ measures (both of which are likely in the power law tail) to fit this model would be inappropriate, because the stochastic differential equation for RGBM (Eq.1) models a lognormal distribution. Therefore, we use $S_{50\%}$, which pertains to an income share (bottom 50%) within the lognormal portion of the distribution, as the appropriate measure to fit our model as described in the algorithm below.

There are two parts to executing the RGBM procedure - first, we estimate drift ($\mu$) and volatility ($\sigma$) of income; and second, we propagate the income dynamics in Eq.1.

We obtain drift ($\mu$) by estimating an exponential fit of the form $\langle x(t) \rangle_N = \langle x(t_0) \rangle_N exp[\mu(t - t_0)]$ for the evolution of mean per capita income over time from $t_0 = 1947$ until $t = 2017$, in 1-year increments. Figure 2 depicts this estimation, yielding $\mu = 0.0231$.

Any proxy used to estimate income volatility ($\sigma$) must ensure that it meaningfully relates to the bulk of the income distribution which comprises a large rural workforce dependent on agriculture, as well as a significant proportion of the urban workforce that works in the informal sector (Naik, 2009; Chand, Srivastava, & Singh, 2017). Long-term time series of granular income data is unavailable for India, but we find very short-term rural daily wage rate time series (2013-19) at monthly granularity, for a number of professions (for both men and women) such as ploughing and tilling, weaving, harvesting and threshing, logging and woodcutting, inland fishing, handicrafts, animal husbandry, horticulture, construction, plant protection, and LMV and tractor driving. We are however able to find much longer time series of a number of possible proxies to capture income volatility, such as wholesale prices of staple crops such as wheat and rice, wholesale price of common commodities like jaggery (gur) as well as the price of gold, which is a common investment in portfolios of most Indian households (RBI, 2017). All of these commodity price series are available at a weekly resolution - wheat, rice, jaggery prices are available for 20 years 1993-2012; and gold prices for 41 years 1979-2019. Data for daily wages was released by the Ministry of Labour and Employment and available from Indiastat (https://www.indiastat.com), crop and commodity prices were obtained from the Open Government Platform of the Government of India (https://data.gov.in/), and gold prices were obtained from World Gold Council

(https://www.gold.org/goldhub/data/price-and-performance). We estimate annualised $\sigma$ for each of these wage and price data sets as the standard deviation of weekly (for commodity price data) or monthly (for rural wage data) logarithmic changes of the prices/wages, multiplied by $(52 \; weeks \; per \; yr)^{0.5}$ or $(12 \; months \; per \; yr)^{0.5}$, based on weekly/monthly resolution of data. These annualised values are averaged to get a consolidated $\sigma$. Using this approach, we compute the following price volatilities: $\sigma(rice) = 0.08, \sigma(jaggery) = 0.13, \sigma(wheat) = 0.14$. For the shorter-term wage time series, we get a range of volatilities: $0.01 \leq \sigma(wage) \leq 0.11$. We also find that volatilities for assets such as gold and the BSE Sensex index tend to be higher: $\sigma(gold) = 0.17, \sigma(sensex) = 0.24$. Therefore, for the purpose of our analysis, we therefore use $\sigma = 0.15$. We also present results for $\sigma = 0.1$ and $\sigma = 0.2$ to assess sensitivity of dynamics to $\sigma$ choice.

Now that we have estimates for both $\mu$ and $\sigma$, our objective is to reproduce the income shares of the bottom 50% ($S_{50\%}$) by fitting a time series $\tau(t)$ - the value of the reallocation parameter over time. The income dynamics of the RGBM algorithm are executed as follows: 1. initialise $N$ individual initial incomes from a lognormal distribution such that the modelled cumulative income of the bottom 50% of the population, $S_{50\%}^{model}(t_0)$, matches the observed value of $S_{50\%}(t_0)$, i.e. $S_{50\%}^{model}(t_0) \cong S_{50\%}(t_0)$; 2. once incomes have been initialized, each of the $N$ individual incomes are propagated using Eq. 1 for $\Delta t = 1$, with the value of $\tau(t)$ chosen to minimize the difference: $abs[S_{50\%}^{model}(t + \Delta t, \tau) - S_{50\%}(t + \Delta t)]$; 3. Step 2 is repeated till the end of the time-series in 2015 to get a full time series for $\tau(t)$. Table 1 lists the model parameters.

| Parameter | Value |
|---|---|
| Drift ($\mu$) | 0.0231 |
| Volatility ($\sigma$) | 0.15 |
| Population ($N$) | 100,000 |

**Table 1**: *Model parameters*

Berman, Peters and Adamou (2017) find that the RGBM yields three distinct regimes of behaviour based on the nature of reallocation (positive, zero, or negative). Using parameters from Table 1 for drift and volatility in the Indian context, we test the model for positive, negative, and no reallocation with $\tau = +0.1; 0.0; -0.1$ respectively (other parameters: $N = 1000$, and $x_i(t_0) = 1 \; for \; i = 1, \dots, N$). For $\tau = 0$ or no reallocation, the RGBM is simply the GBM, which does not converge to a stationary distribution in the long-time limit, and both mean income and income inequality increase continually over time (Figure 3a). Mean income shows growing divergence from the median over time. For $\tau > 0$, or positive reallocation, which we expect to reflect the reality of most modern economies which have systems of taxation and redistribution, incomes disperse (which means that inequality may still increase despite redistribution) but remain confined around an increasing sample mean $\langle x \rangle_N$ (Figure 3b). The median of the distribution also rises over and remains close to the mean. As $\tau$ increases – indicating increasing redistribution from top to the bottom - the distribution is more closely held around the mean. For $\tau < 0$ or negative reallocation, income is essentially redistributed from the poor to the rich (Figure 3c). In a regime where $\tau < 0$ over time, incomes diverge from the mean, and the reduction of incomes at the bottom directly contributes to growth of incomes of those at the top of the distribution. There is no stationary distribution as incomes diverge exponentially away from the mean.

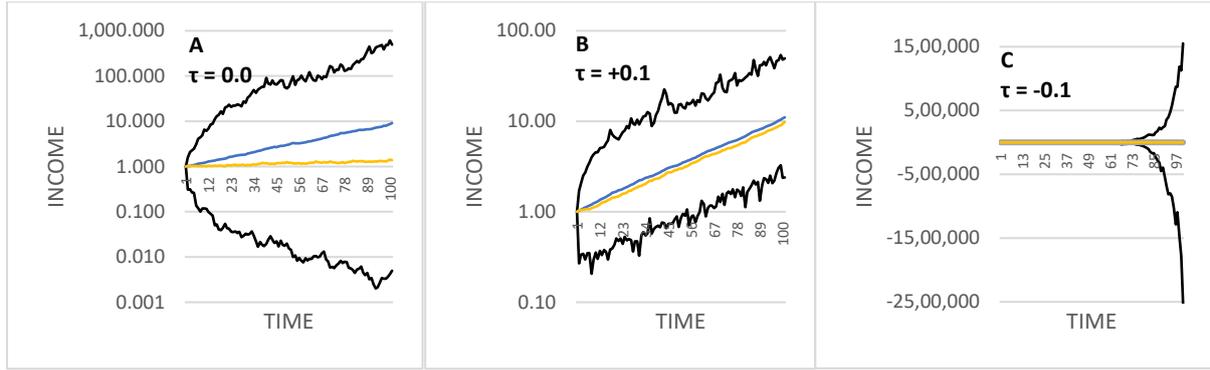

**Figure 3**: *Simulated reallocation regimes.* A: Zero reallocation ($\tau = 0.0$). This is simply the GBM where incomes follow a lognormal distribution. Mean income and income inequality increase over time. B: Positive reallocation ($\tau = 0.1$). Incomes disperse but stay around sample mean for finite $\tau$. Mean income increases with time. C: Negative reallocation ($\tau = -0.1$). Incomes diverge exponentially from the mean, no stationary distribution exists and redistribution occurs from the bottom to top of distribution. Black lines: Maximum and minimum incomes forming the income envelope. Blue line: Sample mean of incomes. Yellow line: Sample median of incomes.

## 3. Temporal evolution of reallocation

We execute the RGBM algorithm as described in Section 2 and find that the simulated income share of the bottom 50% of the population (Figure 4a, dotted green) is in close accordance with the empirical data (Figure 4a, solid blue) over the entire time period under consideration. This close correspondence in fitting $S_{50\%}$ is obtained by appropriate choice of the reallocation parameter time series (termed $\tau_{50\%}$). The temporal evolution of $\tau_{50\%}$ reveals that reallocation is largely positive between 1951 and 2002, and then persistently negative for a decade, before showing a rise again (Figure 4b, solid blue). It could be reasonably argued that the reallocation described by $\tau_{50\%}(t)$ in Figure 4b shows too much variability on an annual basis and that reallocation policies in an economy cannot possibly result in such sharp changes year on year. In order to address this and smooth the evolution of $\tau(t)$, we compute an effective reallocation rate (termed $\tilde{\tau}_{50\%}$) as the 5-year moving average of $\tau_{50\%}$ - i.e. at given time $t = t_y$, the effective reallocation rate $\tilde{\tau}_{50\%}(t_y)$ is the simple average of the reallocation rates at times $t_y, t_{y-1}, .., t_{y-4}$. In order to verify that the effective reallocation rate $\tilde{\tau}_{50\%}(t)$ is still representative of the same income distribution as the simple reallocation rate $\tau_{50\%}$ and not introducing any other systematic element, we use $\tilde{\tau}_{50\%}(t)$ to propagate the initial income distribution and compute the resultant $S_{50\%}^{model}(\tilde{\tau}_{50\%}, t)$. Figure 4a (dashed red) plots the temporal evolution of $S_{50\%}^{model}(\tilde{\tau}_{50\%}, t)$ and we see that it shows close alignment with $S_{50\%}(t)$. Therefore, the effective reallocation rate appears to be a meaningful measure of the actual redistribution occurring in the income distribution.

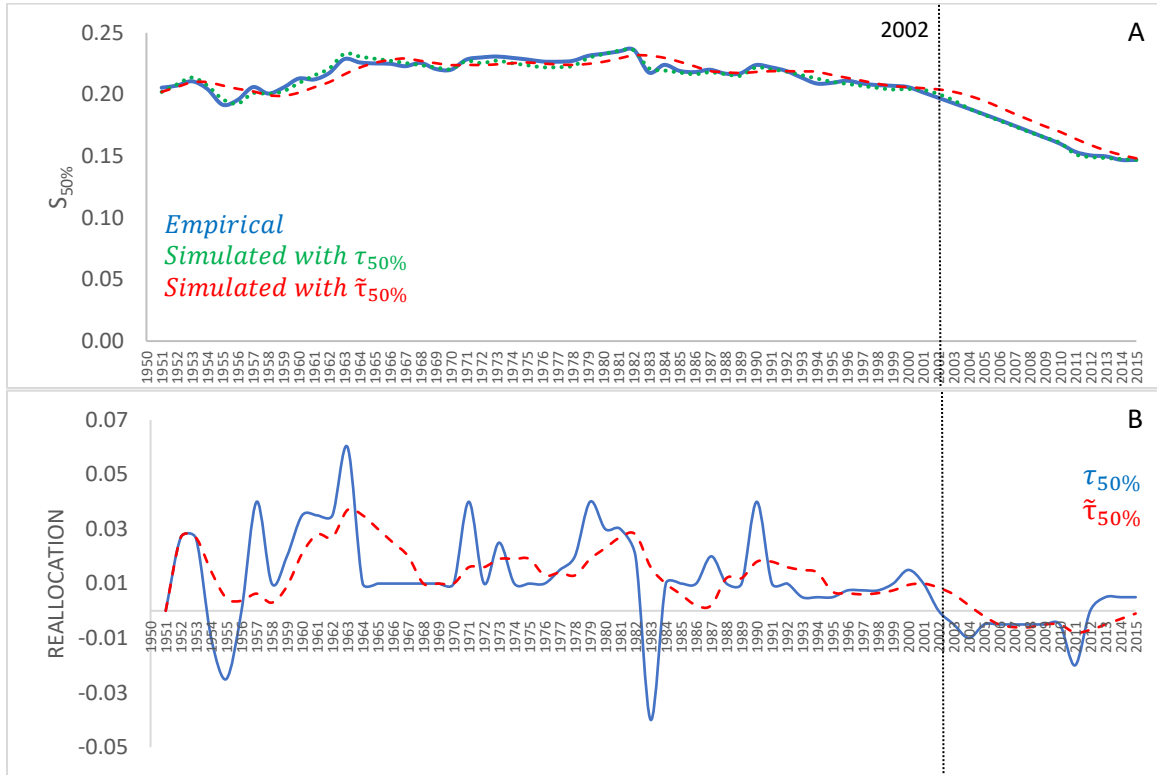

**Figure 4:** *Temporal evolution of income inequality and reallocation (1951-2015).* A: Fraction of income belonging to the bottom 50% of population over time. Solid blue line: Actual Fraction of income owned by bottom 50% of population ($S_{50\%}$). Dotted green line: Simulated fraction of income owned by bottom 50% of population as per reallocation rate $\tau_{50\%}(t)$. Dashed red line: Simulated fraction of income owned by bottom 50% of population as per effective reallocation rate, $\tilde{\tau}_{50\%}(t)$. The curve described by using $\tilde{\tau}_{50\%}$ is closely aligned with the actual data, and the effective reallocation rate is therefore a meaningful measure of actual redistribution. B: Reallocation over time. Blue line: Reallocation Rate ($\tau_{50\%}$) over time. Dashed red line: Effective Reallocation Rate ($\tilde{\tau}_{50\%}$) over time. Effective reallocation rates $\tilde{\tau}_{50\%}(t)$ are positive for close to five decades from 1951 and then become negative from 2005. The dotted black line represents the year 2002, at which point the reallocation rate enters the negative regime and the income share of the bottom half begins showing a sharp decline.

Figure 4b (dashed red) shows the evolution of $\tilde{\tau}_{50\%}$, revealing that that income inequality essentially entered a new and persistent regime of negative reallocation in the mid-2000s, though a declining trend in reallocation is apparent since the 1980s. In order to verify the sensitivity of this result, we use the RGBM model to also estimate the effective reallocation rates $\tilde{\tau}_{50\%}(t)$ for $\sigma = 0.1$ and $\sigma = 0.2$ as well. In general, we see that the evolution of effective reallocation rate in all scenarios follows similar trends of rise and decline over time, with the only salient difference being the levels of the curves – the curve for $\sigma = 0.2$ is the higher (Figure 5a, dashed blue) and that for $\sigma = 0.1$ is lower (Figure 5a, dashed red) than our base case of $\sigma = 0.15$. In all cases, we find a declining trend of reallocation from the mid-1980s, and even for $\sigma = 0.2$, reallocation drops to zero in the mid-2000s (Figure 5a). Overall, this suggests that the regime of negative reallocation observed post 2002 in our base case is a robust result.

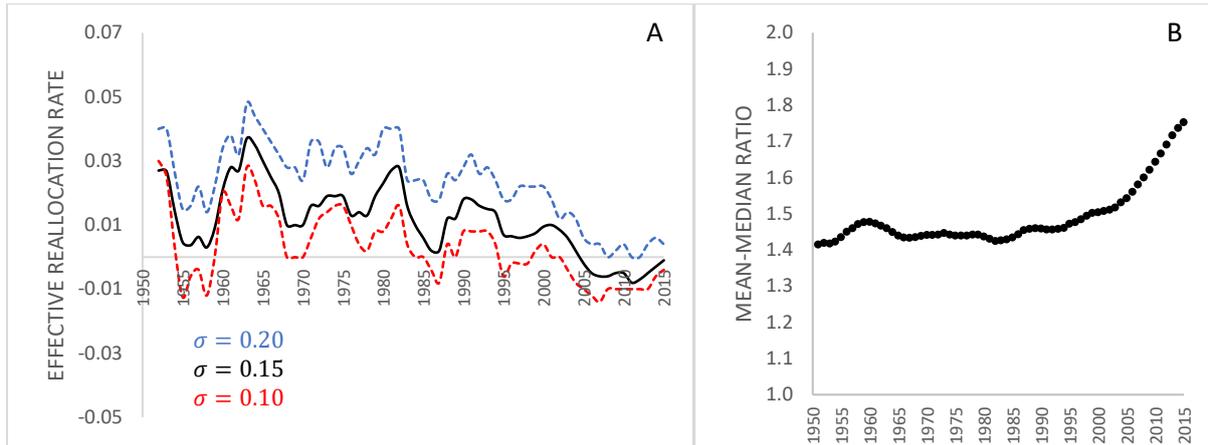

**Figure 5:** *Sensitivity of effective reallocation and central tendencies of income distribution.* A: Evolution of effective reallocation rates ($\tilde{\tau}_{50\%}$) for different levels of income volatility. Black line: $\tilde{\tau}_{50\%}$ for $\sigma = 0.15$. Dashed red line: $\tilde{\tau}_{50\%}$ for $\sigma = 0.1$. Dashed blue line: $\tilde{\tau}_{50\%}$ for $\sigma = 0.2$. All paths show similar trends, just at different levels. B: Mean-Median ratio of rescaled income distribution (1951-2015). Incomes diverge and the mean pulls away from the median in the early 2000s.

## 4. Implications of negative reallocation

The transition from a positive to negative reallocation regime coincides with the onset of a steep 27.5% drop in the share of the bottom half of the population between 2002 and 2015 (Figure 4). This drop in income share of the bottom half is compatible with the following possibilities since the early 2000s: (a) that relative rates of income growth are, on average, higher towards the top of the distribution, though all parts of the distribution experience non-negative growth; or (b) that there are declines in real income lower in the distribution accompanied by growth in income higher in the distribution, indicative of a regressive transfer of resources from poor to rich that is generating negative reallocation. We explore the change in income shares and the growth rates at different parts of the distribution to precisely interpret the meaning of the observed negative reallocation in the rescaled income distribution.

When we assess the evolution of mean and median of the rescaled income distribution, we find that both measures grew through 1951 to 2015, with the mean always higher than the median. However, when we look at the ratio of mean to median income over time, we find that it declines from 1.48 to 1.42 between 1960 and 1983, but then it steadily grows to 1.5 by 2000, before steeply rising to 1.75 by 2015, highlighting the divergent nature of the income distribution in recent times (Figure 5b). We elicit further proof of this phenomenon by tracking the evolution of income shares of each decile of the population over the period 1951 to 2015 (Figure 6a). Until 1983, we see evidence for progressive redistribution (income convergence) with income shares of the top income deciles decreasing and those of the bottom deciles increasing. As we move forward in time from 1983, we find that the bottom deciles own a decreasing share of income, and post 2002 this rate of decline in income share worsens. From a peak income share of 2.7% in 1983, the bottom decile (Decile 1) sees this decline to 2.1% in 2002, and then rapidly to 1% in 2015 (the corresponding shares for the bottom percentile – Percentile 1 - are 0.18%, 0.13%, and 0.03%) (Figure 6b). We test the robustness of this result for the bottom decile by varying income volatility ($\sigma = 0.1$ and $\sigma =$

0.2) and find that the extent of decrease in income share across these scenarios is in close agreement with the base case - income share of bottom decile in 2015 is 0.76% for $\sigma = 0.1$ and 1.16% for $\sigma = 0.2$, compared to 1% for base case (Figure 6b).

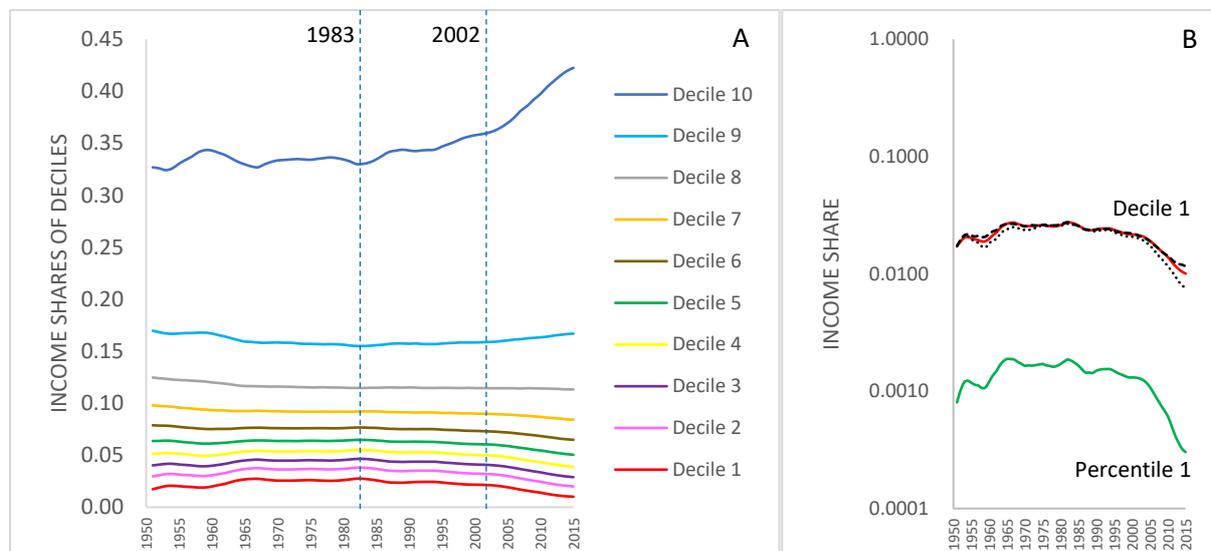

**Figure 6.** *Temporal evolution of income shares by decile from 1951-2015.* A: Share of income earned by each decile over time. We see mild convergence in incomes till 1983, and then a gradual divergence, which is heightened post 2002. Income share of the bottom decile shows the sharpest decline. B: Sensitivity of income shares to income volatility. Outcomes of income of bottom decile appear robust to assumptions about income volatility. Solid red line: Bottom decile (base case). Dotted black line: Bottom decile ($\sigma = 0.1$). Dashed black line: Bottom decile ($\sigma = 0.2$). Solid green line: Income share of bottom percentile (base case).

It is important to point out that the rise in income share of the top decile (Decile 10) is underestimated here (actual share of top decile in 2015 is 56%, as against 42% from the model), because the model does not account for the power law operating at the tail of the income distribution as discussed previously. This means that the income shares of the middle 40% (Deciles 2 – 5) are overestimated in our model. However, given that we fit the model based on income earned by the bottom 50%, our outcomes for that part of the distribution remain consistent. Overall, the decline in income shares is apparent across each of the bottom 5 deciles of the rescaled income distribution - though at increasing rates as we go the lower in the distribution (for instance, compare the bottom decile and bottom percentile in Figure 6b). Therefore, while the extent of redistribution estimated by our model is conservative, the nature (direction) of such redistribution remains robust to tail (rich) incomes in the income distribution.

Next, we estimate the growth incidence curves for the rescaled income distribution for each decade from 1951 to explore temporal evolution of income growth in different parts of the distribution (Fig. 7). We have already seen that the income inequality declines until 1983, with increasing income shares for the lower parts of the distribution (Fig. 6a). This is complemented by the growth incidence curves for 1951-60 and 1961-70, which show a progressive decline in growth rates as we go up higher in the rescaled income distribution, clearly indicating that the higher rates of growth lower in the distribution were also driving convergence in incomes over this time (Fig. 7a). The growth incidence curves after 1980 – 1981-90 and 1991-2000 - describe a different regime, with increasing growth rates higher in the distribution, corresponding to an increase in income inequality and a declining but positive reallocation parameter ($\tilde{\tau}_{50\%}$). This is an indication that even though there is a rise in income inequality as evinced by declining income shares and comparatively lower (but positive) growth rates at the bottom of the distribution, the nature of reallocation implied in the distribution still retains a progressive character with resources being transferred from the rich to the poor.

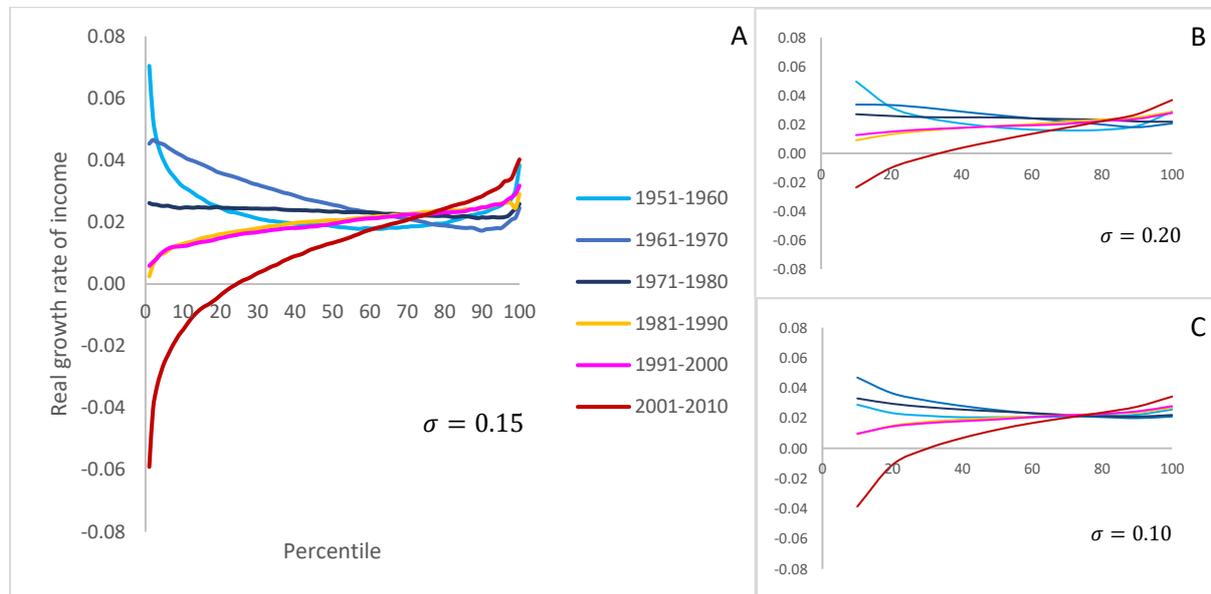

**Figure 7.** *Growth incidence curves by decade*. A: Growth Incidence Curve by percentile for $\sigma = 0.15$. Growth of incomes at different points in the distribution transitions from a progressive to regressive regime, with bottom deciles showing negative income growth for 2001-10. B: Growth Incidence Curve by decile for $\sigma = .20$. C: Growth Incidence Curve by decile for $\sigma = 0.10$. Like in the base case, we see that income growth for the bottom deciles is negative in the 2000s indicating that negative reallocation is a robust result.

However, in the decade 2001-10, the nature of the income distribution appears fundamentally altered with both declining income shares and declines in real income lower in the distribution (the bottom two deciles) when compared to higher growth rates higher in the distribution. It is in this regime that we observe the emergence and persistence of negative reallocation ($\tilde{\tau}_{50\%} < 0$) over many years. This is indicative of a diverging income distribution, which is a significant concern given that it means that income redistribution has left a progressive regime (reallocation from the rich to the poor) and entered a perverse regressive regime where incomes of the poor are being redistributed to the rich. According to our results, between 2001-10, the income of the bottom decile declined by 2.5% and that of the bottom of percentile declined by 5.9% per annum, while the incomes of the top decile and percentile increased by 3.5% and 4% per annum respectively (Fig. 7a). We find that declining incomes at the bottom of the distribution between 2001-10 occur for all choices of $\sigma$ (bottom decile sees income growth of -2.5% and -3.9% for $\sigma = 0.2$ and $\sigma = 0.1$ respectively), indicating the robustness of this outcome (Figs. 7b, 7c).

In order to empirically validate our findings, we use the data pertaining to the lowest part of the distribution from the income distribution of Chancel and Piketty (2019). Given the methodological concerns they have highlighted in generating this data, we assess the complete set of 54 scenarios used to construct the Indian income distribution. These scenarios arose out of different assumptions for combinations of four critical variables, namely: saving profiles of lower consumption groups (A0: base case with possibility of negative savings rate for the poor, A1: savings profile from IHDS dataset, and A2: no negative savings rate among the poor); choice of survey for estimation of income (B1: IHDS dataset, B2: NSSO dataset); level of distribution up to which survey data is reliable and beyond which tax data is reliable (C1: up to 90$^{th}$ percentile, C2: up to 95$^{th}$ percentile, C3: up to 80$^{th}$ percentile); and strategy for progression of income levels and thresholds at a given time (D1: convex junction profile, D2: linear profile, D3: concave profile) (Chancel & Piketty, 2019). We compute rate of growth of average income at the bottom of the distribution from 2001-10 for all 54 possible combinations and find that just the 6 variations of A-B combinations encompass the entire set of possible outcomes from our computations; that is, for a given A-B combination, all combinations of C and D yield the same results. We find that the rate of average income growth from 2001-10 for the bottom ventile and the bottom decile are negative in 3 out of the 6 scenarios, and for the bottom percentile in 4 scenarios (Fig. 8). For instance, in the A0-B2 scenario, growth rates of average income in the bottom decile, bottom ventile, and bottom percentile were -12.36% (-1.45% per annum), -12.06% (-1.42% per annum), and – 12.09% (-1.42% per annum) respectively between 2001 and 2010.

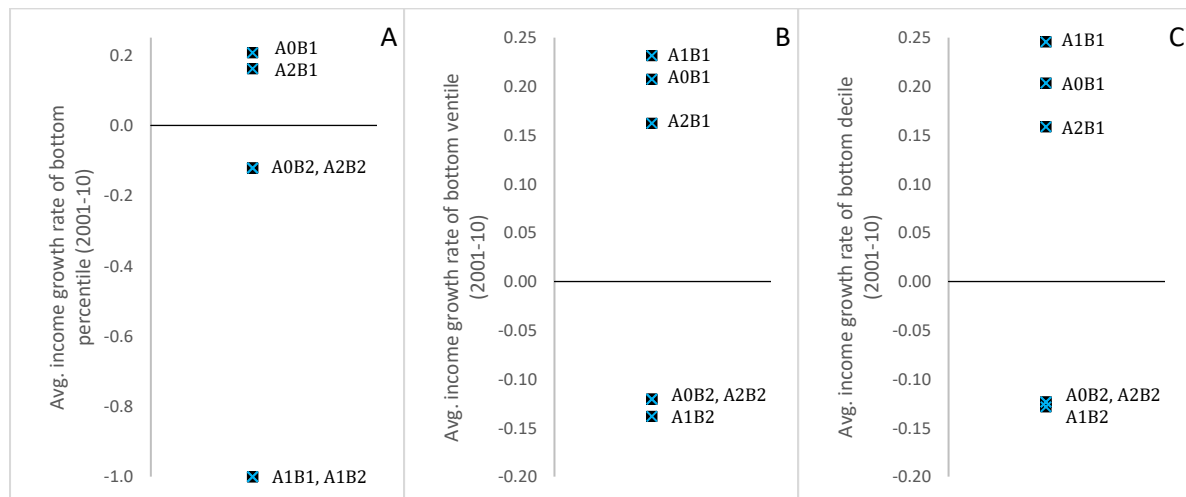

**Figure 8.** *Empirical observations of real average growth 2001-10*. A: Average real income growth rate of bottom percentile (2001-10). Income growth is negative in 4 scenarios – A0B2, A1B2, A2B2, and A1B1. B: Average real income growth rate of bottom ventile (2001-10). Income growth is negative in 3 scenarios – A0B2, A1B2, and A2B2. C: Average real income growth rate of bottom decile (2001-10). Income growth is negative in 3 scenarios – A0B2, A1B2, and A2B2.

This empirical test confirms that in at least half the scenarios under consideration a significant proportion of the bottom of the income distribution displays negative real income growth (and not just declining income shares), as predicted by the prolonged period of negative $\tau$ post 2002 in our model. That our modelled outcome is quite possibly a reflection of persistent regressive reallocation in the Indian income distribution post 2000 raises fundamental questions about the nature of economic growth and redistribution in India, and specifically its impact on the most economically vulnerable populations in the country.

**4. Discussion**

It is well recognized fact that economic growth is essential for a nation like India to effectively combat poverty (Roemer & Gugerty, 1997; Fosu, 2017; Adams Jr., 2004; Planning Commission, 1962). Economic growth has been seen as key to the reduction of poverty in India over the past 40 years (Deaton & Dreze, 2002; Dhongde, 2007; Panagariya & More, 2014; Panagariya & Mukim, 2014), and recognizing that the increased growth may indeed be somewhat inequitably distributed, it is argued that the benefits of growth are still spread across the income distribution, leaving individuals better off than before (Bhagwati & Panagariya, 2013). Kuznets (1955) argued that some level of inequality was inevitable as economic growth happened and that redistribution would follow economic growth, though there is evidence to suggest that lower inequality benefits economic growth and therefore poverty reduction (Fosu, 2017; Lakner, Mahler, Negre, & Prydz, 2019; Alesina & Rodrik, 1994). Our finding that for the past two decades India has been, and perhaps continue to be, in a regressive regime of negative reallocation, where inequality is not just increasing, but that there is a degenerate redistribution of income from the bottom to the top, underlines the need for a deeper interrogation into the nature of economic growth in India.

Prior work has studied the fall and rise of economic inequality in India in the context of structural economic conditions, developments in the political economy, and global economic changes (Banerjee & Piketty, 2005; Chancel & Piketty, 2019; Deaton & Dreze, 2002; Kohli, 2012; Dev & Ravi, 2007). It is recognized that from 1947, one of the explicit goals of the mixed economy under Jawaharlal Nehru was the curbing of elite economic power, and the declining share of income of the top 1% (and top 10%) till the early 1980s is found to be consistent with the role of socialist policies - such as state ownership of the 'commanding heights' of the economy, price regulations, import barriers, and progressive tax structures (with very high top marginal rates) - in driving convergence in the income distribution (Banerjee & Piketty, 2005; Chancel & Piketty, 2019). Since the early 1980s under Rajiv Gandhi and especially in the 1990s under Narasimha Rao and subsequent governments, there was a move away from socialism towards economic liberalisation, incorporating a set of policies including trade openness, price deregulation, increase in imports, tax reduction (especially of top marginal rates), and denationalisation of industry, that resulted in sharp increases both in economic growth and income inequality (Banerjee & Piketty, 2005; Kohli, 2012; Chancel & Piketty, 2019; Rodrik & Subramanian, 2004; Basole, 2014). Our work suggests that we have been in a regime of negative redistribution since the early 2000s, and it is plausible the implementation of the National Rural Employment Guarantee Act (NREGA) in 2005 by the Manmohan Singh government (Ministry of Rural Development, 2005) has had some impact in redressing the extent of this perverse redistribution. While many challenges in its implementation are acknowledged, the NREGA program is found have yielded higher incomes, higher political participation amongst disadvantaged groups, and improved labour force participation especially among women in rural India (Bhatia & Drèze, 2006; Shankar & Gaiha, 2013; Azam, 2011; Freud, 2015). The income impact of this program on the rural workforce since 2005 could be one possible explanation for the decrease in magnitude of negative redistribution by 2015 (although overall redistribution still remains negative). But, the NREGA program appears to have been significantly diluted, restricted, and underfunded by the NDA government since 2014 (Freud, 2015; Bhalla, 2014), possibly enhancing the risk of India remaining in the regime of negative income redistribution for longer. Income data post 2014 will be required to confirm subsequent redistribution trends.

The fundamental structural changes in the Indian economy from the 1980s onward also echo broader global trends, which have resulted in the reversal of gains in income inequality in nations across the world (Alvaredo, Chancel, Piketty, Saez, & Zucman, 2017; Chancel & Piketty, 2019; Assouad, Chancel, & Morgan, 2018). This continues even today with the onset of the high technology revolution in the first two decades of the 2000s, which has led to a renewed rise in global inequality (Milanovic, 2016), and India appears no exception to this trend (Chancel & Piketty, 2019; Deaton & Dreze, 2002; Sarkar & Mehta, 2010). While it has meant excess returns for capital (Milanovic, 2016) (there were nine billionaires in India in 2000, 57 in 2011, and 131 in 2017 as per the Forbes Rich List), it has at the same time resulted in the increasing informalization of jobs in the organized sector (Mehrotra, Gandhi, Saha, & Sahoo, 2012; NSSO, 2015). This is obvious not only in the nature of employment in new-age technology companies (such as Uber, Ola, Swiggy, and Amazon) which provide their services through networks of agents who are not directly employed by them (McQuown, 2016), but also in the increasingly contractual nature of employment in the manufacturing sector - where the fraction of contractual employees increased from 16% in 1998-99 to 35% in 2014-15 (Mehrotra, Gandhi, Saha, & Sahoo, 2012; NSSO, 2015). The continual casualization of the workforce in the formal sector has meant a gradual stripping away of job contracts, security and benefits, resulting in diminished possibilities for meaningful worker mobilization and organization (Applebaum & Lichtenstein, 2016).

Nowhere is the stark nature of the India's extant income distribution more apparent than in the agricultural sector, which employs close to 50% of India's workforce (Dept. of Economic Affairs, 2018). It is well recognized that agrarian economic distress has been widespread in India since the 1990s (Vakulabharanam & Motiram, 2011; Vaidyanathan, 2006; Reddy & Mishra, 2008). This is manifested in the increasing indebtedness of farmers - over half the nation's farmers are indebted, and both incidence and extent of indebtedness have been growing over time (Suri, 2006; Narayanamoorthy & Kalamkar, 2005) - primarily on account of rising input costs due to removal of public subsidies, output price volatility, and decline in public investments (Vakulabharanam & Motiram, 2011; Suri, 2006; Reddy & Mishra, 2008). This indebtedness is linked to the increase in the number of marginal and small-hold farmers, and to the spate of farmer suicides - over 298,000 farmers committed suicide between 1995 and 2012 (Kennedy & King, 2014; Nagaraj, Sainath, Rukmani, & Gopinath, 2014; Suri, 2006; Vaidyanathan, 2006). Given that casual labourers, smallholders and marginal farmers comprise the bottom of the income distribution (Sinha, Pearson, Kadekodi, & Gregory, 2017), our findings that incomes in the bottom decile and bottom percentile are in continuous and sharp decline since the early 2000s corresponds with the broader evidence on deep agrarian distress. If the state of negative redistribution ($\tau < 0$) persists over time, it is possible that we may even see the emergence of negative incomes at the bottom of the distribution. There is previous evidence of negative income observations, as in the case of Taiwan in the 1960s and 1970s (Pyatt, Chen, & Fei, 1980), resulting from household financial losses in agriculture or micro and small businesses, because of the absence of any distinction between 'household income' and 'business income' in these informal contexts (Chen, Tsaur, & Rhai, 1982). Consequently, if the current negative redistribution trend holds, there is real concern that increasing fractions of the workforce at the bottom of India's income distribution will be net debtors in the economic system.

Given this confluence of global and national trends, there is a need for structural interventions to enable a reversal of the extreme inequality evident today. In designing responses to address this situation, it is important to remember that India still remains a low-income country and that there is a need for both continued economic growth as well as inequality reduction. A starting point for this would be a recognition that the current Indian economic model is leaving a substantial proportion of the population disconnected from the growth process – as our work reveals, the bottom-most deciles see their income share shrink since the early 2000s. Indeed, almost this exact recognition of the limitation of economic growth processes was explicated in the Planning Commission's assessment of the development process in newly independent India – they estimated that about 20% of the population remained outside of economic development processes at the time (in 1961), and would need specific policies to secure their basic economic well-being (Planning Commission, 1962). Therefore, the salient question, both then and now, is how policy can ensure growth for the largest possible proportion of the population while simultaneously enabling meaningful support and redistribution to systematically benefit those left behind, so as to reduce inequality.

It has been argued that the current tide of rising income inequality can be countered by a number of strategies such as new forms of political mobilization, taxation policies, universal incomes, and sustained increase in public investments (Piketty, 2014; Schiller, 2004; Skidelsky & Skidelsky, 2012; Burman, 2014; Glomm & Ravikumar, 2003; Subramanian, Belli, & Kawachi, 2002). Schiller (2004), for instance, contends that a progressive tax system is an essential bulwark against income inequality, by ensuring that higher earnings are taxed at higher rates, but that the system does not respond adequately if income inequality rises and become increasingly more extreme (as our work demonstrates in the Indian context). To counter such inequality, he proposes a fundamental reform of the tax system by having taxes indexed to income inequality. This would mean that the system remains progressive, but most importantly, tax rates would endogenously adjust to changes in inequality. Effectively, in scenarios of increasing or extreme inequality, the rate of rise in marginal tax rate on the highest income brackets will reflect the rate of rise in inequality. Burman (2014) further nuances this idea by proposing a progressive tax code integrating inequality indexing with inflation indexing, where losses in tax revenue on account of inflation indexing can be offset by increased tax revenues from inequality indexing. The nature of this offset due to inequality indexing would be that richer tax payers bear more of the burden (and poorer tax payers less) in case of worsening inequality. Piketty (2014) also recommends raising the tax rates on the highest incomes, as well as increasing inheritance taxes. The Universal Basic Income (UBI), where all citizens of a country receive a regular, unconditional sum of money from the government, is proposed as a counter to inequality. It is argued that since the lion's share of productivity gains over the past few decades have gone to the richest, a reversal of this trend could fund a modest initial basic income (Skidelsky & Skidelsky, 2012). Given the increasing threat of automation and the further exacerbation of inequality that this trend could represent, a UBI that grows in line with capital productivity would benefit the many instead of privileging the few (Skidelsky & Skidelsky, 2012). Significantly increasing public investments in education and health so that a reasonable quality of these services is accessible to every last citizen is potentially another way to ensure improved long-term redistribution outcomes (Subramanian, Belli, & Kawachi, 2002; Glomm & Ravikumar, 2003).

While the details of specific policy proposals to counter income inequality will need deeper consideration, what our work specifically highlights is the need for structural reforms to ensure that the divergence of the income distribution is reversed and we are able to return to a persistently progressive redistribution regime.

## 5. Conclusion

We attempt to characterize the dynamics of redistribution in the Indian income distribution. Milanovic (2016) shows that the rise of modern capitalism after the Industrial Revolution had a fundamental impact on the nature of the relationship between average income and income inequality. Essentially, both average income and income inequality rise over time and it is only for a brief period in the mid-20th century that we see a decline in income inequality even as average income increases. The advent of the communications and internet revolutions has once again resulted in a continual upward surge in inequality over the past three decades. In keeping with global trends, we find that income inequality in India reduces between 1951 and the early 1980s, beyond which it shows continual increase, which is particularly sharp post 2002.

Given this empirical characterization of the evolution of income post the Industrial revolution, we seek to model income evolution as a multiplicative process following Geometric Brownian Motion (GBM). In doing so, we follow the reallocating GBM (RGBM) methodology of Berman, Peters, and Adamou (2017), where they incorporate a redistribution parameter $\tau$ to the GBM to capture the direction and magnitude of reallocation occurring within the distribution.

Applying the RGBM to Indian income inequality data, we find that there are two distinct redistribution regimes: between 1951 and 2002 the reallocation is positive ($\tau > 0$) and between 2002 and 2015, the reallocation is distinctly negative ($\tau < 0$). Model outcomes suggest that while the entire bottom half of the income distribution has seen a shrinking share of income in this time, those at the very bottom have been worst hit, with the bottom decile earning just ~1% (and the bottom percentile just ~0.03%) of national income in 2015. Using growth incidence curves, we also find that since the early 2000s, the bottom deciles of the income distribution have experienced declines in real income (with bottom decile experiencing -2.5% growth per annum), while the top deciles experience the highest growth rates within the distribution. We also empirically verify these model results against the scenarios used by Chancel and Piketty (2019) to construct the Indian distribution, and find that in at least half of the scenarios considered, the bottom decile, bottom ventile, and bottom percentile see negative average real income growth from 2001-10, thus providing considerable support to model outcomes. Therefore, the emergence and persistence of negative $\tau$ in our model appears indicative of the reality of perverse, regressive reallocation, where resources are being redistributed from the poor to the rich in India's income distribution.

We discuss how the nature of India's economic growth is closely linked to the negative redistribution apparent in the income distribution. The increasing informalization of the formal workforce in both new-age technology as well as traditional manufacturing sectors has meant that workers have been left with no avenues for mobilization. The agricultural workforce appears to be the worst hit, with highly volatile incomes, especially of casual labourers, marginal and small farmers, combining with rising indebtedness resulting in increased impoverishment over time. We argue that the current model of economic development leaves out a significant proportion of the population, and that meaningful responses to the situation must consider the need for both high economic growth and lower income inequality in India. We discuss a number of ways to counter inequality such as inequality indexed taxes, increased public investments in health and education, and the Universal Basic Income. Overall, it is important for us to reconsider and suitably reorient extant models of economic growth so that prosperity is more equitably distributed and any economic re-distribution remains progressive.

Our work has limitations. We also do not model the dynamics of generating the upper tail of incomes, which follows a power law, because of our focus on the lognormal portion of the distribution. However, incorporating this would provide a more complete model for the entire income distribution. Further work using the RGBM model on alternate measures of income inequality such as the Gini coefficient time series for India could help further validate our results.